\documentclass[twocolumn,english,aps,prl,longbibliography]{revtex4-1}
\usepackage[T1]{fontenc}
\usepackage[latin9]{inputenc}
\usepackage{color}
\usepackage{amsmath}
\usepackage{graphicx}
\usepackage{natbib}
\usepackage{braket}
\usepackage[normalem]{ulem}

\usepackage[colorlinks=true,urlcolor=blue,citecolor=blue,linkcolor=blue]{hyperref}

\makeatother

\usepackage{babel}

\newcommand{\fogi}{$\ket{f0}\leftrightarrow\ket{g1}~$}

\begin{document}

\title{Pulsed reset protocol for fixed-frequency superconducting qubits}

\author{D. J. Egger, M. Werninghaus, M. Ganzhorn, G. Salis, A. Fuhrer, P. M\"uller, S. Filipp}
\affiliation{IBM Research GmbH, Zurich Research Laboratory, S\"aumerstrasse 4, 8803 R\"uschlikon, Switzerland}

\begin{abstract}
Improving coherence times of quantum bits is a fundamental challenge in the field of quantum computing. With long-lived qubits it becomes, however, inefficient to wait until the qubits have relaxed to their ground state after completion of an experiment. Moreover, for error-correction schemes it is important to rapidly re-initialize syndrome qubits. We present a simple  pulsed qubit reset protocol based on a two-pulse sequence. A first pulse transfers the excited state population to a higher excited qubit state and a second pulse into a lossy environment provided by a low-Q transmission line resonator, which is also used for qubit readout. We show that the remaining excited state population can be suppressed to $1.7\pm 0.1\%$ and that this figure may be reduced by further improving the pulse calibration.
\end{abstract}

\date{\today}

\maketitle

\section{Introduction}

Building a fully operational and useful quantum computer requires many  good, high-coherence qubits along with high-fidelity gates. An equally important requirement is the ability to (re-)initialize qubits  in their ground state \cite{DiVincenzo2000}. In the context of error correction codes \cite{Fowler2012}, the reset of syndrome qubits must be accomplished on timescales much shorter than the coherence time, at best comparable to the duration of single-qubit pulses. Another field of application is speeding up the execution of variational quantum algorithms  \cite{Barrett2013,Peruzzo2014,OMalley2016,McClean2016,Eichler2015,Kandala2017} in which the quantum system has to be re-initialized quickly to start the next gate sequence with modified parameters. An efficient reset mechanism can also be used for cooling qubits \cite{Valenzuela2006}.

The easiest way to reset a qubit involves waiting several times the decay ($T_1$) time so that the energy stored in the qubit relaxes into the environment. As the $T_1$ times of superconducting qubits increase to values above $100~\mu\rm{s}$ \cite{Rigetti2012} this passive reset becomes inefficient and requires millisecond waiting times. Several methods have been developed to actively reset superconducting qubits. One approach is to (re-)initialize the qubits by measuring their individual states and inverting them via a $\pi$-pulse conditioned on the measurement outcome \cite{Riste2012, Govia2015}. Such an active reset protocol suffers from the relatively long latency times, typically in the few hundred nanoseconds range, of the readout-chain and state descrimination. It is also limited by the fidelity of the qubit measurement. An un-conditional reset protocol with no need for active feedback, as proposed here, requires less time and does not rely on fast hardware components. Such a mechanism requires a low-temperature dissipative environment that is coupled to the qubit in a controlled way. A superconducting qubit may, for instance, be coupled to a low-temperature resistive bath by controlling a coupling resonator \cite{Tuorila2017}. In a circuit QED setting \cite{Blais2004,Wallraff2004} the dissipative environment may be provided by a transmission line resonator with low quality factor that may also be a readout resonator. Frequency-tunable qubits can be rapidly tuned into resonance \cite{Reed2010}, a scheme that is also used for generating traveling single-photon Fock-states \cite{Bozyigit2011,Eichler2012}. When the qubit frequency cannot be tuned in-situ \cite{Chow2011} and the coupling between qubits and other circuits is fixed, coherent microwave pulses provide the only way to manipulate the system. In such an architecture fast qubit initialization has to be implemented either with a quantum-circuit refrigerator \cite{Tan2017} or with microwaves. It has been demonstrated that the system can be steered by multi-tone microwave drives into a steady-state that leaves the qubit in the ground state \cite{Geerlings2013}. Recently, in a paper also including a review of other reset methods, a fast unconditional reset protocol was demonstrated in an architecture using two low quality factor resonators, one for qubit readout and one for qubit reset \cite{Magnard2018}.

Here, we demonstrate that an effective Jaynes-Cummings type interaction from a single external drive can successfully reset the qubit within a few hundred nanoseconds using only the readout resonator. This avoids introducing a second resonator used for reset and thus saves space on the chip. Similarly to \cite{Magnard2018}, our reset protocol is based on an induced Rabi oscillation between the second-excited state of a superconducting transmon-type qubit and a harmonic oscillator \cite{Zeytinoglu2015}. This technique has been employed to generate shaped microwave photons \cite{Pechal2014} to establish entanglement between remote qubits \cite{Narla2016, Kurpiers2017, Campagne-Ibarcq2018}. It has also been used to measure a vacuum-induced geometric phase \cite{Gasparinetti2016}, a quantized version of Berry's well-known geometric phase. An attractive feature of this tunable coupling mechanism is its applicability to fixed-frequency qubit architectures, as for instance used in the IBM Q Experience \cite{QuantumExperience}. It can even serve to entangle qubits by generating a holonomic transformation in a three-dimensional subspace spanned by the resonator and two qubits \cite{Egger2018}. 
\section{Description of the system \label{Sec:setup}}
The system is made up of a fixed-frequency superconducting qubit coupled with strength $g$ to a co-planar waveguide resonator used both for read-out and reset of the qubit.  The Hamiltonian describing the system \cite{Pechal2014}, in a frame rotating at the frequency $\omega_\text{d}$ of the drive with a slowly-varying complex envelope $\Omega(t)$, is
\begin{align} \notag
\hat{H} =&\, \delta_\text{r}\hat a^\dagger\hat a+\delta_{\rm q} \hat b^\dagger\hat b^{\phantom{\dagger}}+\frac{\alpha}{2}\hat b^\dagger\hat b^\dagger\hat b\,\hat b  \\
+& g\left(\hat b^\dagger\hat a+\hat b\,\hat a^\dagger\right)+\frac{1}{2}\left(\Omega(t)\hat b^\dagger+\Omega^*(t)\hat b\right). \label{Eqn:H}
\end{align}
Here $\hbar=1$, $\hat b$ ($\hat b^\dagger$) is the qubit lowering (raising) operator whilst $\hat a$ ($\hat a^\dagger$) is the resonator lowering (raising) operator. 
The effective transition energies between the qubit and resonator are $\delta_\text{q}=\omega_\text{ge}-\omega_\text{d}$ and $\delta_\text{q}=\omega_\text{r}-\omega_\text{d}$, respectively.
The qubit has a transition frequency $\omega_\text{ge}/2\pi=4.904~\rm{GHz}$ between its ground state $\ket{g}$ and its first excited state $\ket{e}$. The anharmonicity is $\alpha/2\pi=-330~\rm{MHz}$ resulting in a transition frequency of $\omega_\text{ef}/2\pi=4.574~\rm{GHz}$ between the first and the second excited state $\ket{f}$. The readout resonator transition frequency is $\omega_\text{r}/2\pi=6.838$ GHz. The qubit resonator coupling $g/2\pi=67~\rm{MHz}$ is computed from the dispersive shift $\chi/2\pi=335\pm48~\rm{kHz}$ measured by probing the resonance of the readout resonator for different qubit states \cite{Koch2007}. The measured $T_1$-times of the qubit and resonator are $48~\mu\rm{s}$ and $235~\rm{ns}$, respectively.

\begin{figure}[!tpb]
\begin{center}
\includegraphics[width=0.48\textwidth]{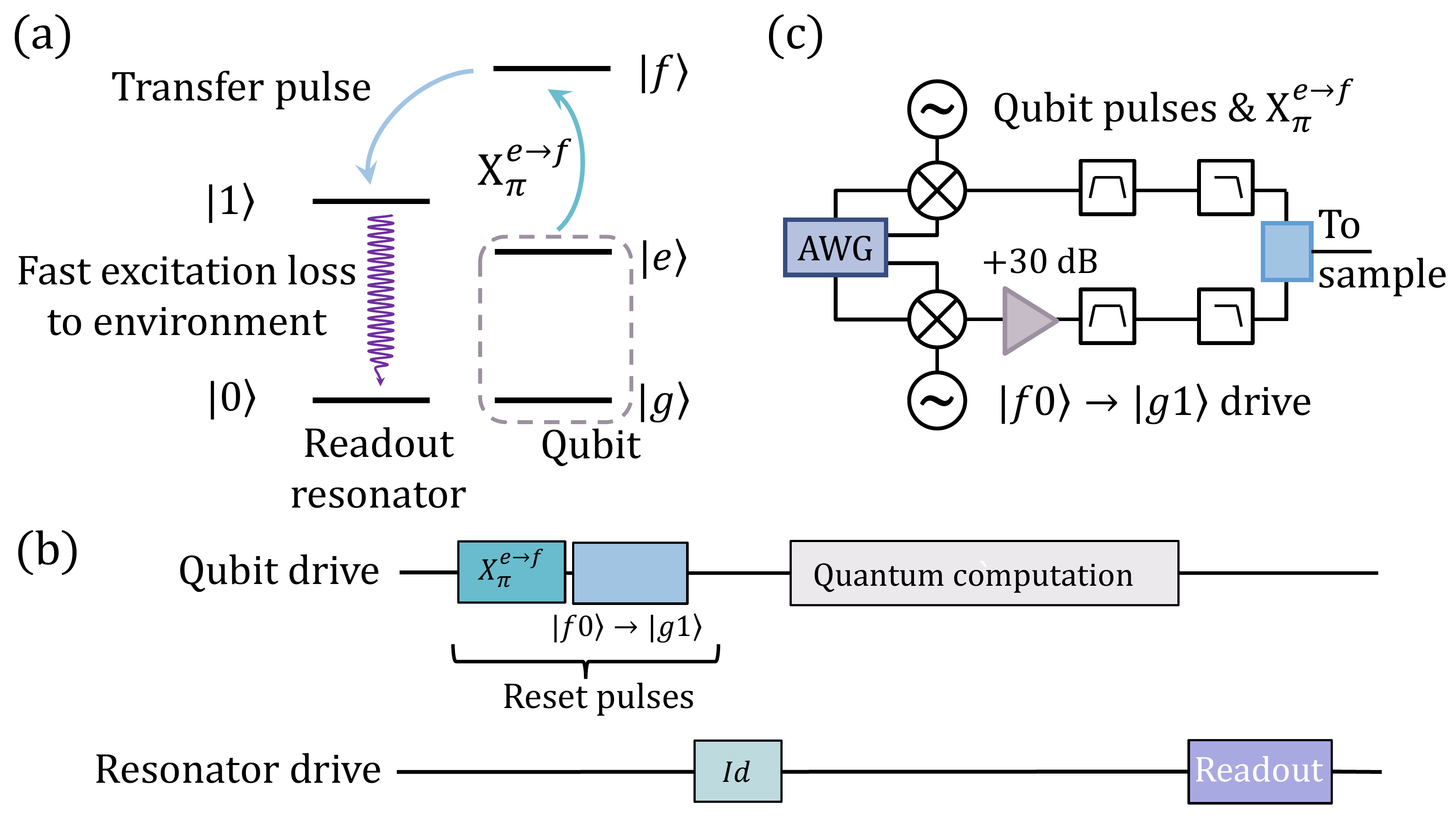}
\caption{\label{Fig:diagrams} (a) Level diagram  of the transmon-resonator system along with the relevant pulses to reset the qubit. (b) Pulsed reset protocol. Before a quantum algorithm commences, the reset pulse sequence is carried out: first the remaining qubit population in state $\ket{e}$ is shelved to state $\ket{f}$ with the pulse $X_\pi^{e\to f}$, second the population in $\ket{f}$ is transferred to the readout resonator with the $\ket{f0}\leftrightarrow\ket{g1}$ drive where it rapidly decays to the environment. (c) Room temperature electronics. A $30~\rm{dB}$ amplifier enhances the rate at which the $\left|f0\right\rangle\leftrightarrow\left|g1\right\rangle$ transition is driven.}
\end{center}
\end{figure}

The protocol described in this paper makes use of the qubit-resonator transitions between the $\ket{f0}$ and $\ket{g1}$ states where $\ket{0}$ is the ground state and $\ket{1}$ is the excited state of the resonator, see Fig.\ \ref{Fig:diagrams}(a). The single-qubit transitions and the qubit-resonator transition  are controlled by applying the real microwave drive $\Omega_0(t) \cos(\omega_{\text{d}} t - \varphi(t))$ with a corresponding complex amplitude $\Omega(t)=\Omega_0(t)e^{i\varphi(t)}$ where $\Omega_0(t)$ and $\varphi(t)$ are slowly varying. Pulse generation details are shown in Fig.~\ref{Fig:diagrams}(c).

Applying a drive at the qubit frequency $\omega_{\text{d}}=\omega_{\rm ge}$ ($\omega_{\rm ef}$) with $\varphi=0$ creates a rotation $X_{\beta}^{g\to e}$ ($X_{\beta}^{e\to f}$) around the x-axis of the $\{\ket{g},\ket{e}\}$ ($\{\ket{e},\ket{f}\}$) Bloch-sphere with angle $\int\Omega(t)\text{d}t=\beta$ \cite{Bianchetti2010}. A different rotation axis in the equatorial plane can be selected by changing the phase $\varphi$ of the drive. Similarly, applying a drive at the difference frequency between the $\ket{f}$ state of the transmon and the excited resonator state $\ket{1}$, i.e. $\omega_{\text{d}}=\omega_{\rm ge} + \omega_{\rm ef}-\omega_\text{r}$, activates induced Jaynes-Cummings-type vacuum-Rabi oscillations between these states. Adiabatic elimination of the qubit $\ket{e}$ state, which is far detuned from the drive frequency gives the effective Hamiltonian \cite{Pechal2014, Zeytinoglu2015}
\begin{align}
\hat H_{\text{eff}}=&\Delta_\text{f0}\ket{f0}\!\bra{f0}
+\tilde{g}\ket{f0}\!\bra{g1}+\text{H.c.} \label{Eq:HeffSingle}
\end{align}
The ac-Stark shift $\Delta_\text{f0}$ is to leading order quadratic in the drive strength $\Omega$. The drive frequency can be set to compensate for this ac-Stark shift so that the states $\ket{f0}$ and $\ket{g1}$ form a degenerate subspace allowing for a coherent population transfer between qubit and resonator. The effective coupling strength between $\ket{f0}$ and $\ket{g1}$ is
\begin{align} \label{Eqn:geff}
\tilde{g}(t)=\frac{g\alpha\Omega(t)}{\sqrt{2}\delta(\delta+\alpha)}
\end{align}
with the qubit-resonator detuning $\delta = \omega_r - \omega_{\rm ge}$. 
The rate of the $\ket{f0}\leftrightarrow\ket{g1}$ microwave activated transition decreases with qubit-resonator detuning but can be compensated by stronger driving with reported effective coupling rates up to approximately $10~\rm{MHz}$ \cite{Zeytinoglu2015}.  

\section{Pulsed Reset Protocol}
\begin{figure}[!tbp]
\begin{center}
\includegraphics[width=0.48\textwidth]{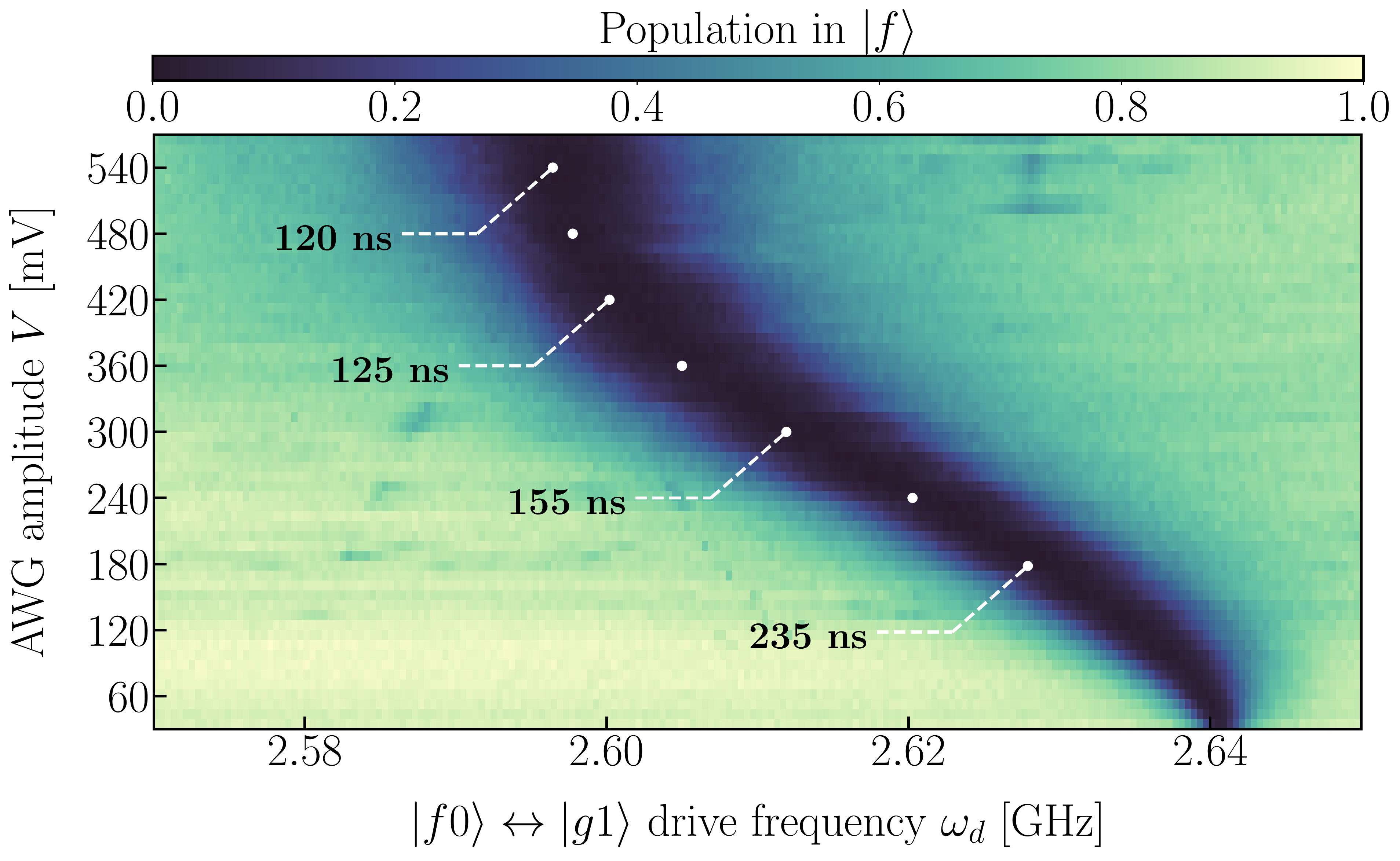}
\caption{\label{Fig:spec} (a) Spectroscopy measurement as a function of the amplitude and frequency of the $\ket{f0}\leftrightarrow\ket{g1}$ drive with the transmon prepared in the $\ket{f}$ state (light color). The drive amplitude is stated as a function of the peak to peak output voltage of the arbitrary waveform generator (AWG). At resonance (dark color), the qubit population is transferred to the low-Q readout resonator. The white dots indicate drive frequency and amplitude pairs for which we performed a time resolved measurement of the population transfer between the qubit and the resonator, see Fig. \ref{Fig:time_rabi}(a). The numbers associated to the white dots indicate the time it takes to transfer the population from the qubit $\ket{f}$ state to the resonator $\ket{1}$ state.}
\end{center}
\end{figure}
We utilize the $\ket{f0}\leftrightarrow\ket{g1}$ controllable quantum link as a fast, simple and unconditional method to reset superconducting qubits using only the readout resonator and an extra microwave drive. Our reset protocol does not require flux-tunable qubits, thus preserving the long coherence times of fixed-frequency transmon qubits. The reset works as follows: before performing quantum computations with the qubit logical states $\ket{g}$ and $\ket{e}$ a $X_\pi^{e\to f}$ pulse transfers any remaining population in state $\ket{e}$ to state $\ket{f}$. We then drive the transition from $\ket{f0}$ to $\ket{g1}$ to transfer the qubit population to the readout resonator. Once in the resonator, the excitation quickly decays to the environment due to the low quality factor of the resonator. The pulse scheme is described in Fig.\ \ref{Fig:diagrams}(b). In this work  each experiment is triggered by a pulse emitted at a rate $R$. We apply the qubit reset immediately after a trigger pulse but before any quantum computations on the qubit are done. An equally valid approach would be to apply the qubit reset after the quantum algorithm (including qubit readout) finishes.

From the measured qubit and resonator frequencies we find the frequency of the $\ket{f0}\leftrightarrow\ket{g1}$ transition at $\omega_{\rm f0g1}/2\pi = (\omega_\text{ge}+\omega_\text{ef}-\omega_\text{r})/2\pi= 2.640~\rm{GHz}$. The exact frequency of this transition  depends on the drive strength because of ac-Stark shifts. To calibrate the drive frequency, we carry out spectroscopy with the qubit initially prepared in the $\ket{f}$ state by applying $X_{\pi}^{g\to e}$ followed by a $X_{\pi}^{e\to f}$ pulse. A $10~\mu\rm{s}$ spectroscopic pulse is applied (at a fixed amplitude and frequency) before the qubit is measured, see Fig.\ \ref{Fig:spec}.
This pulse sequence is repeated for different amplitudes and frequencies. 
Figure~\ref{Fig:spec} does not exhibit Rabi oscillations as function of drive amplitude since the applied drive is much longer than the resonator $T_1$ time.
We observe a shift of the resonance frequency obtained from a Lorentzian fit to the spectroscopy curves for each amplitude value. At small drive powers, this resonance line can be fitted to a second order polynomial in drive amplitude. 
The maximum rate at which we can transfer the population from the transmon to the resonator is limited by the amplifier. We reach the 3dB compression point of the amplifier when the output voltage of the AWG is maximum. The non-linearity of the amplifier explains the non-quadratic behavior of the ac-Stark shift at the higher AWG amplitudes seen in Fig.~\ref{Fig:spec}.
Simultaneously driving the qubit $\ket{e}\leftrightarrow\ket{f}$ transition and the qubit-resonator $\ket{f0}\leftrightarrow\ket{g1}$ transition may make the reset sequence more compact but could complicate the calibration of the ac-Stark shifts \cite{Magnard2018}.

\begin{figure}[!tbp]
\begin{center}
\includegraphics[width=0.48\textwidth]{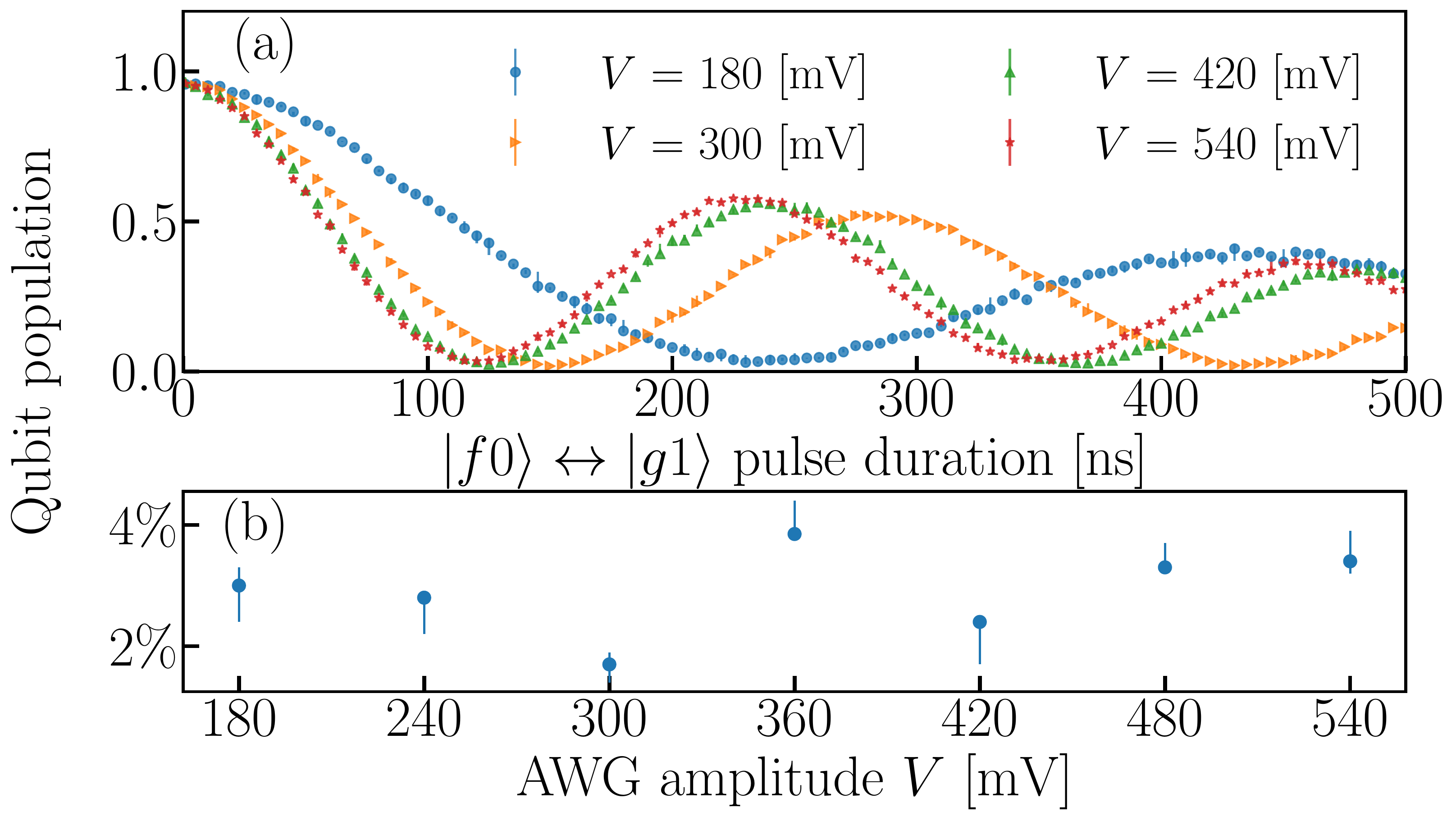}
\caption{\label{Fig:time_rabi} (a) Time resolved Rabi oscillations between $\ket{f0}$ and $\ket{g1}$ for different AWG output voltages indicated by the white dots in Fig.~\ref{Fig:spec}. The transmon is initially prepared in state $\ket{f}$ using an $X_\pi^{g\to e}$ followed by a $X_\pi^{e\to f}$ pulse. (b) Population remaining in the qubit at the first minimum shown in (a). In both (a) and (b) each point is the median of five measurements and the error bars are the 25 and 75 percentiles.}
\end{center}
\end{figure}

\begin{figure*}[!t]
\begin{center}
\includegraphics[width=\textwidth]{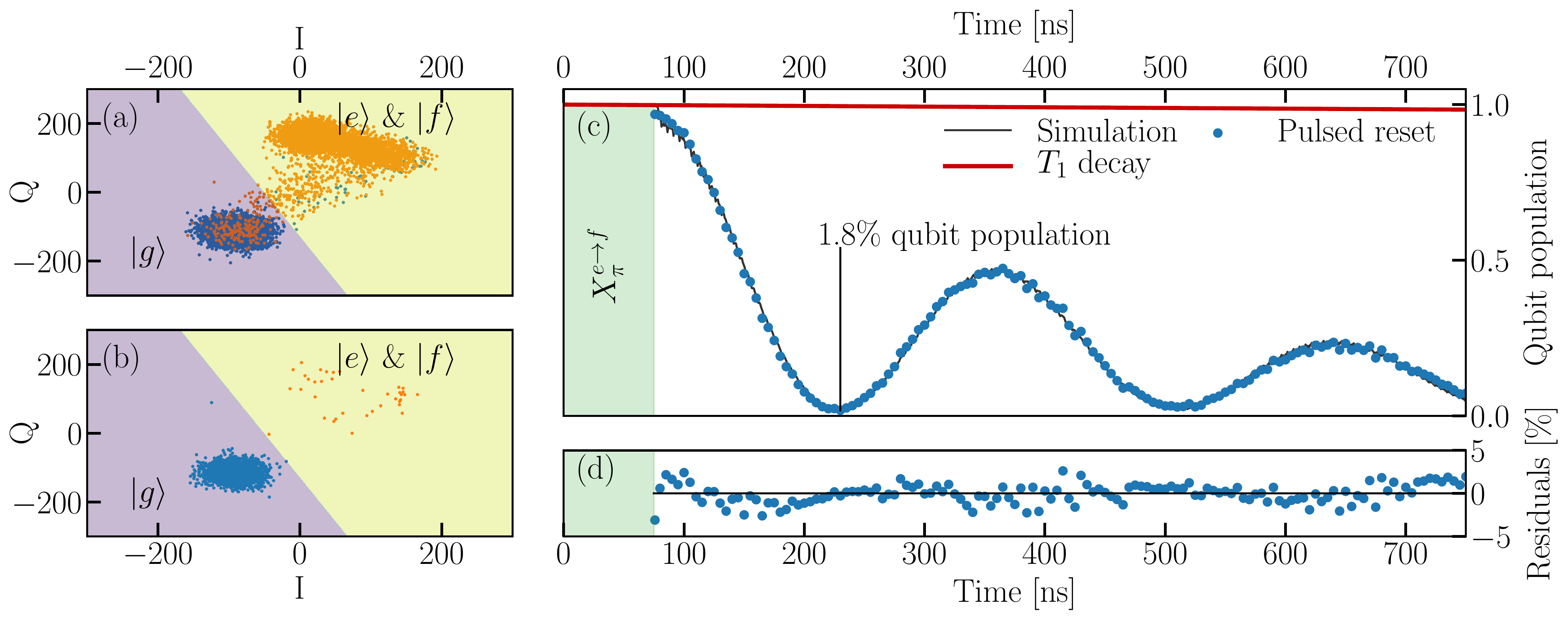}
\caption{\label{Fig:time_trace}
(a) Single shot calibration data using the $\ket{g}$ and $\ket{f}$ states to train a linear support vector machine classifier. The resulting decision boundary is indicated by the background color. The blue and orange dots correspond to the training data knowing that no $\pi$-pulse and a $\pi$-pulse, respectively, have been applied to the qubit before measuring. (b) Single shot data after a $155~\rm{ns}$ \fogi pulse. The population in the qubit is 1.8\% as determined from 2000 single shot measurements. The decision frontier obtained from (a) was used to distinguish whether each data point corresponded to the qubit being in an excited or the ground state. (c) Time resolved trace of the transmon population comparing the pulsed reset to the natural $T_1$ decay (thick red line) of the qubit. The qubit is initialized in the $\ket{e}$ state with a $\ket{g}\to\ket{e}$ $\pi$-pulse. The pulsed reset time trace is measured starting after the $\ket{e}\to\ket{f}$ $\pi$-pulse which lasts $75~\rm{ns}$ whereas the $T_1$-time trace starts immediately. The simulated time trace (thin black line) is obtained from a master equation. (d) The difference between the measured qubit population in (c) and the solution of the master equation describing our system.}
\end{center}
\end{figure*}

In a series of time-resolved measurements at different AWG voltages and $\ket{f0}\leftrightarrow\ket{g1}$ drive power pairs, indicated by the white dots in Fig.~\ref{Fig:spec}, we observe  Rabi oscillations as a function of \fogi pulse duration, see Fig.\ \ref{Fig:time_rabi}(a). With a measured resonator decay rate $\kappa_\text{r} = 4.26~(\mu\rm{s})^{-1}$ similar to the measured Rabi rates ranging from $4.17~(\mu\rm{s})^{-1}$ to $2.13~(\mu\rm{s})^{-1}$, damped induced vacuum-Rabi oscillations are observed indicating that only a fraction of the population returns to the qubit. The optimal time for the reset pulse for each drive amplitude is given by the point where the transfer probability of the qubit excitation to the resonator is maximized. At this point we measure the residual population of the qubit which ranges from $1.7\%$ to $3.9\%$ and shows no systematic dependence on the drive strength, see Fig.~\ref{Fig:time_rabi}(b), but which we attribute to errors in the pulse calibrations. 

The population in the transmon is determined using single-shot measurement data based on the $\ket{g}$ and $\ket{f}$ states used as calibration. Each single-shot amplitude and phase of the $10~\mu\rm{s}$ long readout pulse is determined in a reflective heterodyne measurement, producing a single point in the I-Q plane. The calibration points are used to train a linear support vector machine classifier, see Fig.~\ref{Fig:time_trace}(a), resulting in an assignment fidelity \cite{Magesan2015}
\begin{align} \notag
\mathcal{F}=1-[P(g\vert f)+P(f\vert g)]/2=98.5\%.
\end{align}
Where $P(x\vert y)$ is the probability of measuring state $\ket{x}$ when the qubit was prepared in state $\ket{y}$. 

To better understand the limitations of the pulsed reset protocol we compare one of the best time traces to a simulated master equation using $\hat H$ from Eq.~(\ref{Eqn:H}), see Fig.\ \ref{Fig:time_trace}(b)-(d). The resonator and transmon are modeled with three and four levels, respectively. The resonator and qubit $T_1$ decay are modeled with the Lindblad operators $\hat L_\text{r}=\sqrt{\kappa_\text{r}}\hat a$ and $\hat L_\text{q}=(T_1^{e\to g})^{-1/2}\ket{g}\!\bra{e}+(T_1^{f\to e})^{-1/2}\ket{e}\!\bra{f}+(T_1^{h\to f})^{-1/2}\ket{f}\!\bra{h}$, respectively. The measured qubit $T_1$ times are $T_1^{e\to g}=44.2\pm 1.8~\mu\rm{s}$ and $T_1^{f\to e}=26.1\pm 1.1~\mu\rm{s}$. The $T_1$ time between the fourth transmon level $\ket{h}$ and the third transmon level is assumed to be $T_1^{h\to f}=T_1^{f\to e}/\sqrt{3}$. The model does not include fitting parameters aside from a small frequency shift added to the $\ket{f0}\leftrightarrow\ket{g1}$ drive to reflect miscalibrations of the ac-Stark shift. In the simulation we find that a $29~\rm{kHz}$ frequency mismatch reduces the residuals to below 2.5\%, see Fig.~\ref{Fig:time_trace}(d). This shift is also responsible for an additional 1.3\% of residual population in the qubit. Without it, the simulation predicts a 0.4\% residual population in the qubit.

\section{Experiments at elevated trigger rates}
The reset protocol can be used to increase the rate at which experiments are carried out by initializing the qubit to its ground state before the start of the next experiment. In normal operation conditions without reset pulse sequences applied, our trigger rate $R$ is usually set to $1~\rm{kHz}$, i.e. $1/R$ is 10-20 times the typical qubit $T_1$-time.

\begin{figure}[!t]
\hfill \includegraphics[width=0.43\textwidth, clip, trim=0 245 0 10]{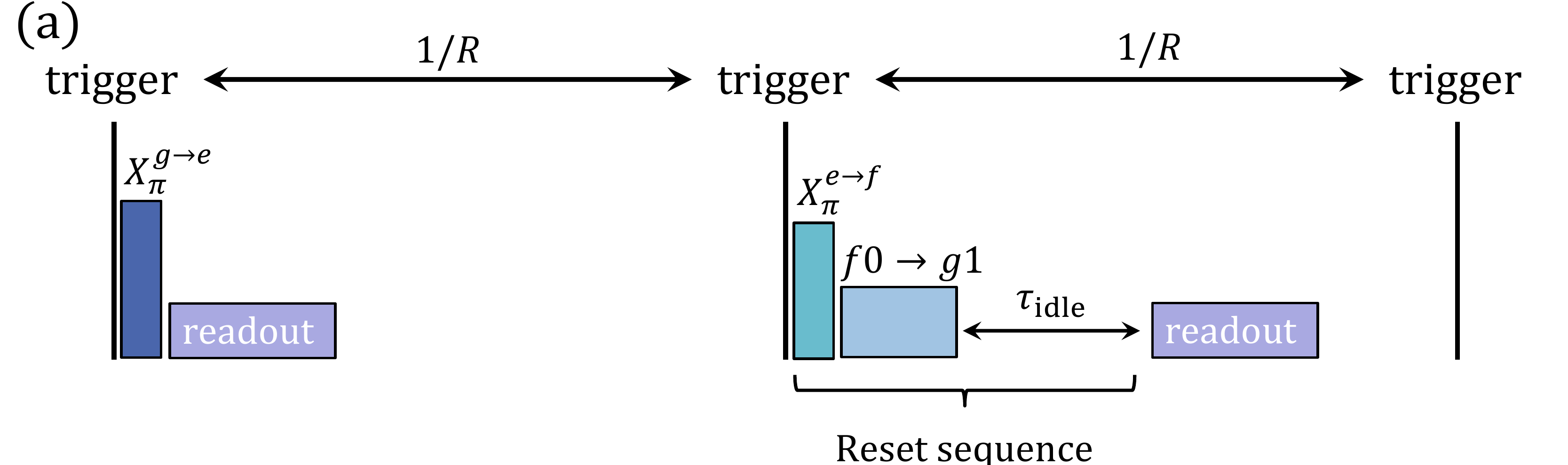}
\includegraphics[width=0.48\textwidth]{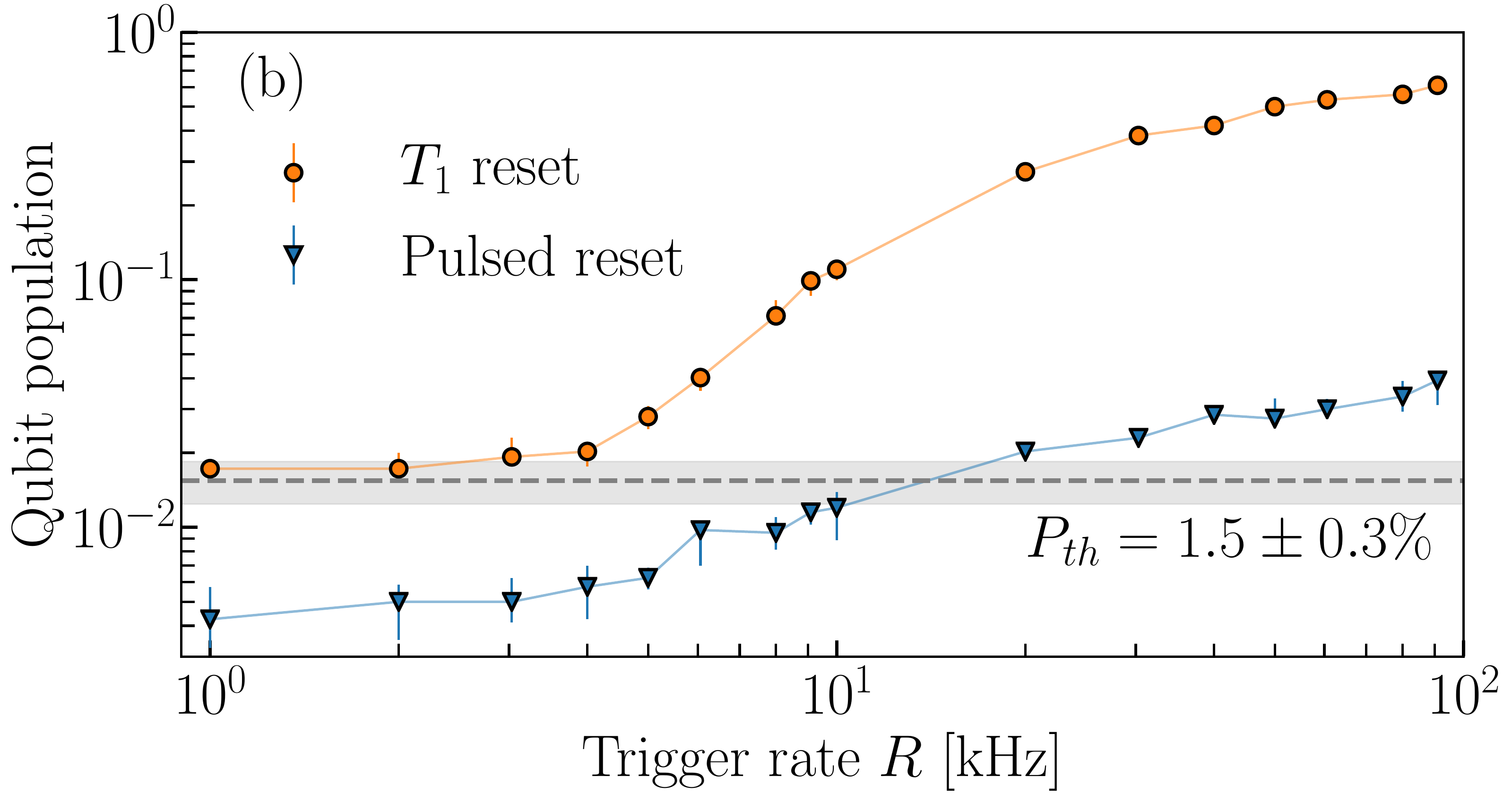} \caption{Population in the qubit for different trigger rates. (a) Pulse sequence. After an initial $X_\pi^{g\to e}$-pulse used to identify $\ket{e}$ in the I-Q readout plane, the qubit is repeatedly measured at each successive trigger pulse. (b) Population in the qubit with (triangles) and without (circles) the pulsed reset sequence. Each data point is the median of ten measurement featuring 2000 rounds. The error bars show the lower 25\% and upper 75\% percentiles of the measured qubit population. The gray dashed line shows that thermal population.
\label{Fig:concept_meas}}
\end{figure}

In this study we prepare the qubit in the exited state $\ket{e}$ with an initial $X_\pi^{g\to e}$ pulse in a first experiment. In a  consecutive experiment after a $1/R$ wait time we only apply the qubit reset pulses and a measurement pulse to determine the qubit population. The $X_\pi^{e\to f}$ pulse lasts $75~\rm{ns}$ while the $\ket{f0}\leftrightarrow\ket{g1}$ pulse lasts $120~\rm{ns}$ (obtained with and AWG output voltage of $V=540~\rm{mV}$). These pulses are followed by a $2~\mu\rm{s}$ idle time. This ensures that the population in the resonator decays before measuring the qubit state. The highest trigger rate used is $R=90~\rm{kHz}$ corresponding to a $1/R=11~\mu\rm{s}$ delay between consecutive experiments. $1/R$ is also the time available for qubit operations including the measurement pulse, which is set to $5~\mu\rm{s}$. Without the reset pulse the remaining population measured in the qubit strongly rises as the trigger rate is increased, see Fig.~\ref{Fig:concept_meas}(b). With the reset pulse, the qubit is emptied as indicated in Fig.\ \ref{Fig:concept_meas}(b) and the residual population only increases slowly as the trigger rate rises. At a rate of $10~\rm{kHz}$ the population in the qubit after the reset pulses is $1.2\%$ which is 9 times better than the $11.0\%$ measured without the reset pulses. As discussed in the previous section, the residual qubit population may be due to miscalibrations in the $X_\pi^{e\to f}$ and $\ket{f0}\leftrightarrow\ket{g1}$ pulses as well as relaxation from $\ket{f}\to\ket{e}$ during the pulse driving the $\ket{f0}\leftrightarrow\ket{g1}$ transition. As the triger rate rises, these errors are made more apparent since the energy relaxation of the qubit has less time to remove any residual population.

\begin{figure}[!t]
\includegraphics[width=0.48\textwidth]{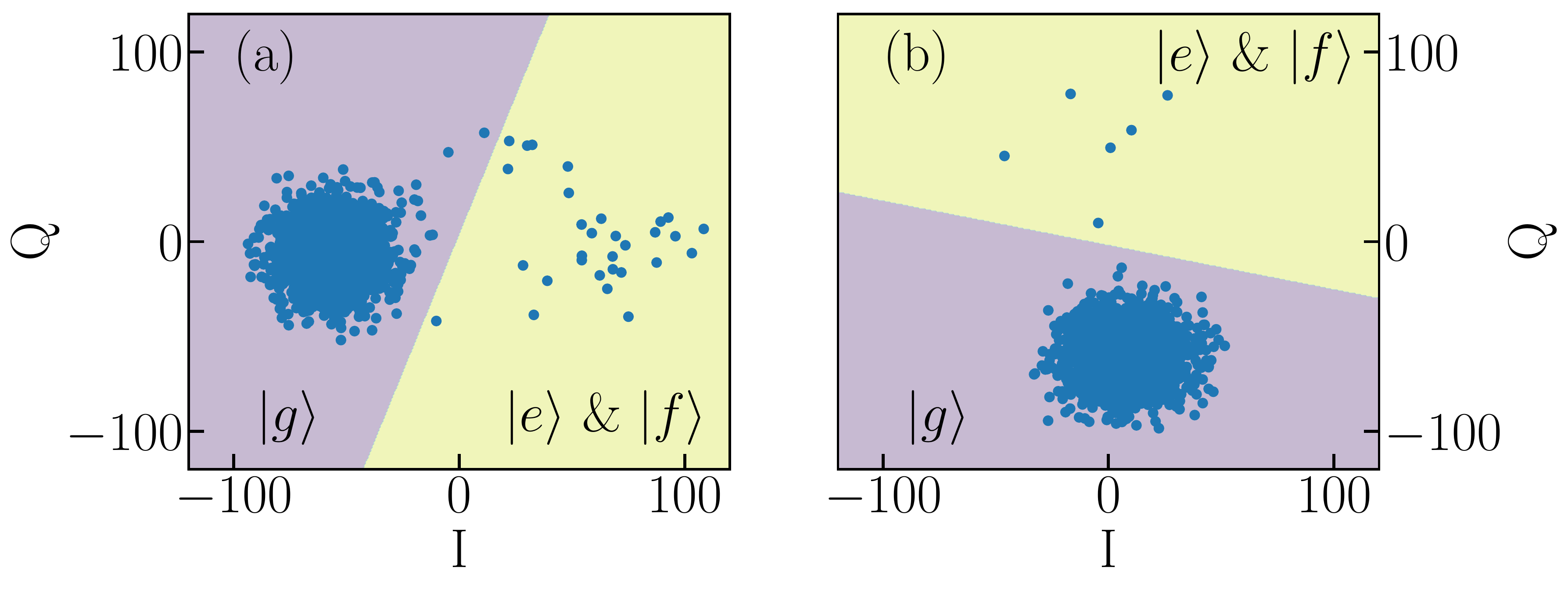} \caption{Single shot data (blue points) in the I-Q plane in the case with $T_1$ reset (a) and with the pulsed reset (b). This data set is one of the ten measurements shown in Fig.~\ref{Fig:concept_meas}(b) at $R=1~\rm{kHz}$. The measured population is $1.73_{-0.10}^{+0.15}\%$ and $0.43_{-0.20}^{+0.16}\%$ with $T_1$ reset and pulsed reset, respectively. The background shows the decision boundary obtained from calibration data (not shown). The dots in the clear and dark regions correspond to the qubit being in the excited states and the ground state, respectively. \label{Fig:cool}}
\end{figure}

We further notice a clear difference in the residual population between the $T_1$ reset and the pulsed reset, even for slow trigger rates. This originates in the active cooling, i.e. initialization, of the qubit by the pulsed reset. This is confirmed by a measurement of the thermal qubit population. We prepare the qubit once in state $\ket{g}$ and once in state $\ket{e}$. 
We then drive Rabi oscillations in the $\{\ket{e}, \ket{f}\}$ subspace by scanning the amplitude of the $X_\pi^{e\to f}$ pulse. 
By comparing the amplitude of the Rabi oscillations when the qubit is initially in $\ket{g}$ to those when the qubit is initially in $\ket{e}$, we find that the thermal population of the qubit $\ket{e}$ state is $1.5\pm0.3\%$, shown as the dashed line in Fig.~\ref{Fig:concept_meas}(b). 
The thermal population matches the population in the qubit at low trigger rates highlighting the limitation of $T_1$ reset, see Fig.~\ref{Fig:concept_meas}(b). At these low trigger rates we see that the pulsed reset protocol cools the qubit by reducing the population to $0.43_{-0.20}^{+0.16}\%$ a much lower value than the thermal population, as evidenced by the single shot data in Fig.~\ref{Fig:cool}(a) and (b).

\section{Conclusion}
We have demonstrated the ability to reset a fixed-frequency transmon qubit within $210~\rm{ns}$ ($2.21~\mu\rm{s}$ when including $8$ times the resonator $T_1$ time) without using complex fast-feedback schemes. The reset protocol relies on a simple square pulse on the \fogi transition that induces vacuum-Rabi oscillations between the qubit and the low quality factor readout resonator. With our reset protocol we are able to increase the  execution speed of our experiment. We also showed that this reset scheme can be used to cool the qubit. We anticipate that the reset pulses may be shaped using methods of optimal control applied to superconducting qubits to compress and improve the reset \cite{Glaser2015, Egger2013a, Egger2014b}. Using a pulse to reset the cavity \cite{McClure2016,Bultink2016}
or a pulse sequence derived from optimal control could decrease the duration of the reset protocol by emptying the readout resonator.

\section{Acknowledgments}
This work was supported by the IARPA LogiQ program under contract W911NF-16-1-0114-FE and the ARO under contract W911NF-14-1-0124. D.E. and S.F. acknowledge support by the Swiss National Science Foundation (SNF, Project 150046).  M. W. acknowledges the support of the European Commission Marie Curie ETN "QuSCo" (GA N$^\circ$765267).

\bibliography{2018_fast_reset}

\end{document}